\documentclass[12pt]{article}

\usepackage[T1]{fontenc}
\usepackage{float}
\usepackage{makecell}
\usepackage{graphicx}
\usepackage[export]{adjustbox}
\usepackage{stackengine}
\usepackage{style/scicite}
\renewcommand\citeleft{[}
\renewcommand\citeright{]}
\makeatletter
\def\@cite#1#2{\leavevmode \cite@adjust
  \citeleft{{#1\if@tempswa\@safe@activesfalse\citemid{#2}\fi
  \spacefactor\@m 
  }}\citeright
 \@restore@auxhandle}
\makeatother
\usepackage{times}  
\usepackage{amsmath}
\usepackage{amssymb}
\usepackage{MnSymbol}
\usepackage[hyphens]{url} 
\usepackage{hyperref}
\usepackage[flushleft]{threeparttable}
\usepackage{color}
\usepackage{colortbl}
\usepackage{booktabs}
\usepackage{multirow}
\usepackage{xcolor}
\usepackage{hyperref}
\usepackage{subfig}
\usepackage{svg}
\usepackage{bm}

\newcommand{\EPINET}{\textsc{SIRNet}}

\title{\EPINET{}: Understanding Social Distancing Measures with Hybrid Neural Network Model for COVID-19 Infectious Spread\\[6pt]
\large \textit{Preprint -- Work In Progress}}

\author{Nicholas Soures$^{\S}$, David Chambers $^{\ddag}$, Zachariah Carmichael$^{\S}$, Anurag Daram$^{\S}$, \\ Dimpy P. Shah$^{\bigcurlywedgedot}$, Kal Clark$^{\bigcurlywedgedot}$, Lloyd Potter$^{\S}$, Dhireesha Kudithipudi$^{\S}$\\
\normalsize{$^{\S}$University of Texas, San Antonio}\\
\normalsize{$^{\ddag}$Southwest Research Institute}\\
\normalsize{$^{\bigcurlywedgedot}$University of Texas Health Science Center, San Antonio}\\
\normalsize{$^{\S}$ $^{\ddag}$ $^{\bigcurlywedgedot}$ MATRIX-AI Consortium}\\
}

\date{}

\begin{document} 

\baselineskip24pt
\maketitle


  \textbf{Abstract:} The SARS-CoV-2 infectious outbreak has rapidly spread across the globe and precipitated varying policies to effectuate physical distancing to ameliorate its impact. In this study, we propose a new hybrid machine learning model, \EPINET{}, for forecasting the spread of the COVID-19 pandemic that couples with the epidemiological models. We use categorized spatiotemporally explicit cellphone mobility data as surrogate markers for physical distancing, along with population weighted density and other local data points. 
  
  We demonstrate at varying geographical granularity that the spectrum of physical distancing options currently being discussed among policy leaders have epidemiologically significant differences in consequences, ranging from viral extinction to near complete population prevalence. The current mobility inflection points vary across geographical regions.
  Experimental results from \EPINET{} establish preliminary bounds on such localized mobility that asymptotically induce containment.
  The model can support in studying non-pharmacological interventions and approaches that minimize societal collateral damage and control mechanisms for an extended period of time. 


\section{Introduction}

Machine learning techniques have offered solutions to many modeling problems, assuming there is abundant data to train a system \cite{big_data_ml}. With the rapid impact of COVID-19, several research groups have begun exploring statistical and mathematical models to study the spread of the disease. One of the early studies using AI \cite{Pneumoni65:online} identified the global spread of the disease through commercial airlines, if the outbreak continues. There are several inefficiencies with the current data available for COVID-19 research, such as limited testing capabilities and high variability within the testing rate (e.g.: 22.08/1000 in Italy, 11.16/1000 in US, to 0.27/1000 in India~\cite{Tounders86:online, Whysomec64:online}), inconsistencies in reporting (under reporting), and publicly available data on infection rates currently are unreliable. Particularly lacking is an understanding of the underlying factors which impact the spread, accuracy and availability of reported cases on a small scale, and quantifiable metrics for how social distancing and quarantine efforts impact the spread. To overcome these challenges in providing early models for forecasting the spread of COVID-19, we combine compartmentalized models with a data-driven machine learning approach.  In doing so, we address a potential pitfall of machine learning (ensuring compliance with the laws of epidemic dynamics) and a limitation of epidemiological models (enabling the creation of complex mappings from available data sources to critical modeling parameters).

In infectious diseases which are human-human transmissible, individual contact rates drive the spread of infectious pathogens across a population over time. To reduce the high contact rate of COVID-19, several efforts have been taken including social distancing and lockdowns. To account for social distancing and lockdown efforts, we use categorized cell phone mobility data from Google LLC~\cite{COVID19C25:online} which includes tracking activity at grocery/pharmacy stores, parks, retail/recreation, residential, train stations, and workplaces globally.  The tracking activity is provided in an aggregated format at the country level, state level, and in the US at the county level. 


This allows us to implement the epidemic model (which can easily be updated to any of the modules), with a dynamic set of parameters which evolve temporally based on social distancing regulations and other non-pharmaceutical interventions to reduce the spread of COVID-19. This is useful as it allows us to model the role of mobility in managing the spread of COVID-19 and provide insight into potential scenarios if restrictions are lifted by different degrees at certain points in time.

\section{SARS-CoV-2}
In December 2019, an atypical case of pneumonia was diagnosed in Wuhan \cite{rodriguez2020clinical,singhal2020review,ralph20202019}, Hubei province of China which was named COVID-19 and the virus was termed as SARS-CoV-24\ \cite{rodriguez2020clinical}. It is a beta coronavirus which is a single-stranded positive sense RNA virus associated SARS-CoV \cite{gorbalenya2020coronaviridae,zou2020single}. Because the disease is relatively new, the epidemiology of the virus has changed significantly so far \cite{sun2020understanding,callaway2020china,guo2020origin,habibzadeh2020novel,adhikari2020epidemiology}. Most cases occur in adults \cite{china2019novel}. It is transmitted from person to person through respiratory droplets (within 6 feet) or less likely through contact with fomites \cite{van2020aerosol,aydin2014influence,yu2016surface}. The average incubation period is 5.2 days, and each case spreads the infection to an estimated 2.2 people \cite{li2020early}. The median duration of viral shedding has been reported by one study as 20 days with range from 8 to 37 days \cite{zhou2020clinical}.

First case of COVID19 in US was diagnosed in state of Washington in January 2020 and now U.S. has the highest toll of infected cases and deaths in the world. With no end in sight for the COVID-19 pandemic, it may continue to spread for next several months leading to millions of infections and thousands of deaths before we reach herd immunity. Herd immunity means when at least 60\% of the population is immune from the infection, either through natural exposure to the virus or vaccine administration. Unfortunately, the current infection/exposure estimates for COVID19 are around 1\% to 2\%.

In the meantime, protective measures such as social distancing, hand washing, avoiding close contact with a sick person, closure of non-essential services, stay at home orders, and covering mouth and nose when around other people are recommended to “flatten the curve” of the epidemic. Without such measures, there can be a high burden on healthcare systems due to the need for hospitalization, intensive care unit admission, and mechanical ventilation.  These mitigation efforts, however, do not come without economic and other human consequences.  Approximately 10 million Americans applied for unemployment benefits in the last 2 weeks of March and world stock markets lost approximately one-third of their value \cite{mckinsey2020}.    

\section{Importance of Mobility in Modeling}
Appropriately tuning mitigation efforts to optimize social welfare has been confounded by the heterogeneous and wave-like pattern of peak-impact globally. The United States, similarly, can expect a high degree of variable impact across metropolitan and micropolitan areas in terms of adherence to policy restrictions on mobility \cite{webster_how_2020}, case fatality rates \cite{rajgor2020many}, seroprevalence \cite{bendavid_covid-19_2020}, and economic impact \cite{wheaton_explore_2020}.  Ultimately policy leaders must, by both necessity and by national policy prescription, translate this variability into an effective reproductive number in order to make optimal policy. The primary challenge posed include the ability to accurately make prescriptive policy in a highly dynamic and geopolitical heterogeneous setting.  The capacity to collect, interpret, and model data to make estimates of current and future effective reproductive numbers is now of paramount concern as the United States undertakes phasic pandemic risk mitigation adjustments in order to optimize population morbidity with economic productivity.  The stakes cannot be higher: where adjustments are made to decrease overall restrictions on mobility, assuming all other factors held constant, it will be met with increases in infection. Several recent studies highlight this behavior \cite{koo2020interventions, kucharski2020early,ferretti2020quantifying}. If the increase in infection is too high, the effective reproductive number will set off a second wave equal to or greater than the initial wave. Similarly, delayed rescindment of restrictions on mobility results in a different type of a second wave: economic disparity, morbidity from inadequately addressed chronic disease, and delayed costs from things like deferred youth education. 

Given the above observations, we believe mobility data offers a unique and powerful opportunity to build a valuable model that can answer these challenges. While mobility is most obviously associated with contact rate we believe it has unrealized value via insight into important public health population characteristics which are also dynamic.  In particular, mobility data may allow us to indirectly incorporate population features such as health literacy, virtual social substitution (the ability of a population to substitute virtual contact for real contact), and business buy-in \cite{webster_how_2020}.  Each of these features may be difficult to measure directly but may be indirectly represented in mobility data.  

The mobility data metrics incorporated into this model succeed on four key population modeling criteria which directly (and indirectly) impact the ability to model effective reproductive number: temporally and spatially explicit coverage, representativeness, and contemporaneousness \cite{deville_dynamic_2014}.  These metrics are collected in a manner common to all regions which a) allows for rapid wide-scale data interpolation and extrapolation and b) accurately reflects the unique geopolitical profile of regions, each with variable laws, customs, socioeconomic profiles, health care resources, and susceptibility rates. Temporally explicit coverage defines the ability of the mobility data to reasonably describe the movement behaviors of people in the three main environmental contexts of a population \emph{(model representation in parentheses)}: work (workplace), non-work non-residential (retail and recreation, grocery and pharmacy, parks, transit stations), and residential (residential). Spatial specificity refers to the ability to collect granular geographical data. Representative refers to the ability for the data to offer a high enough sample size to reasonably reflect the population statistic.  Contemporaneousness reflects that population statistics are dynamic; the ability of the model to adjust and incorporate changes in behaviors in near real-time permits the model to be prescriptive and responsive to data aggregated across regions.  Integration of these metrics into a usable public health resource necessitates a novel and robust modeling strategy.

\section{Proposed Hybrid Model: \EPINET{}}
In this research, we focus on learning and forecasting the
trends in time series via a hybrid model of neural networks and epidemiological models. The forecasting network, referred to as \EPINET{} (named after the foundational epidemiological model), learns from i) a sequence of prior trends that carry long-term contextual information (global time-series) and ii) more recent data inputs that are raw (local time-series) and can inform the forecasting of any abrupt changes.



\begin{figure}[H]
    \centering
    \subfloat[]{\includegraphics[width=0.49\linewidth]{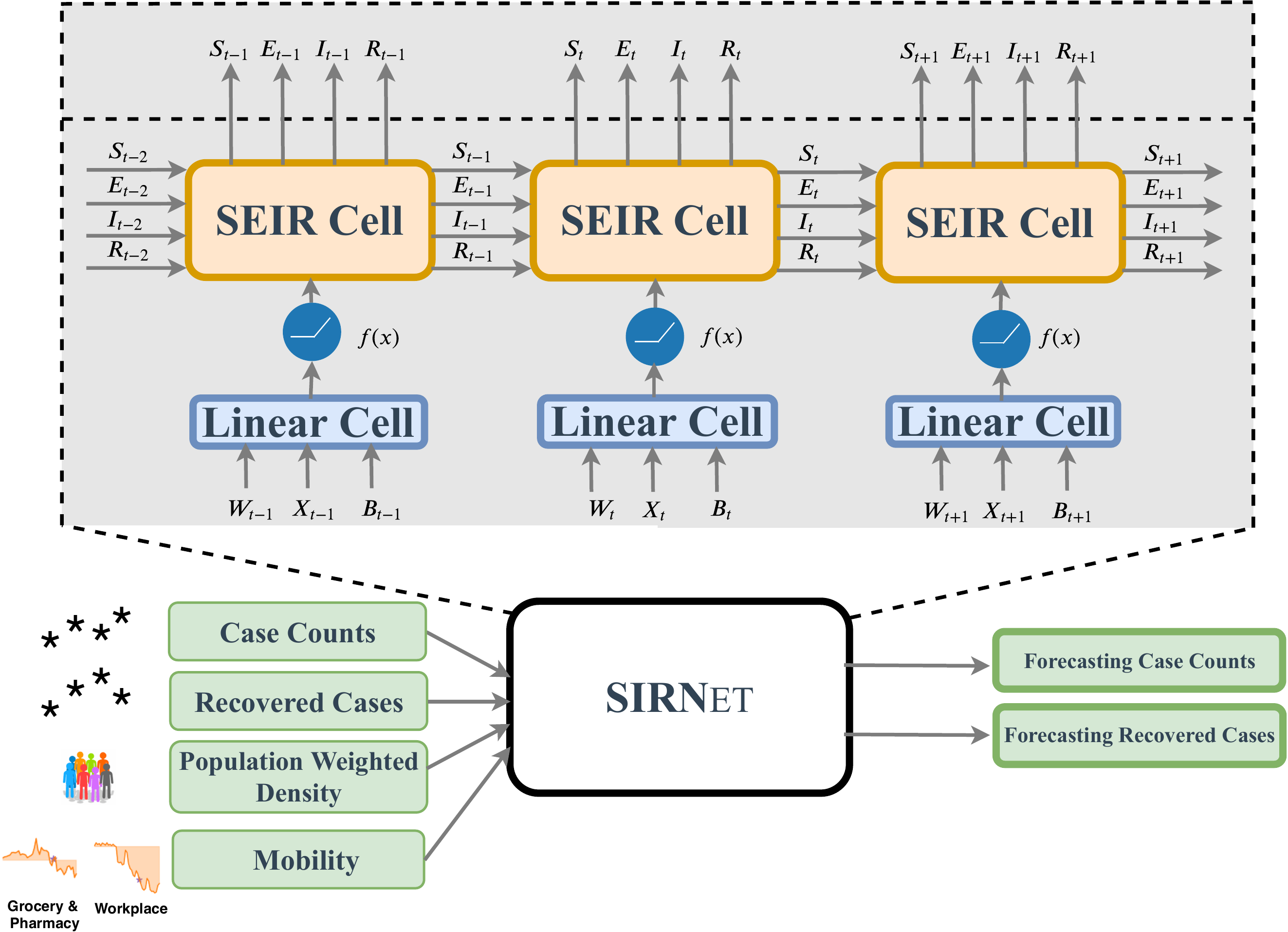}}
    \subfloat[]{\includegraphics[width=0.49\linewidth]{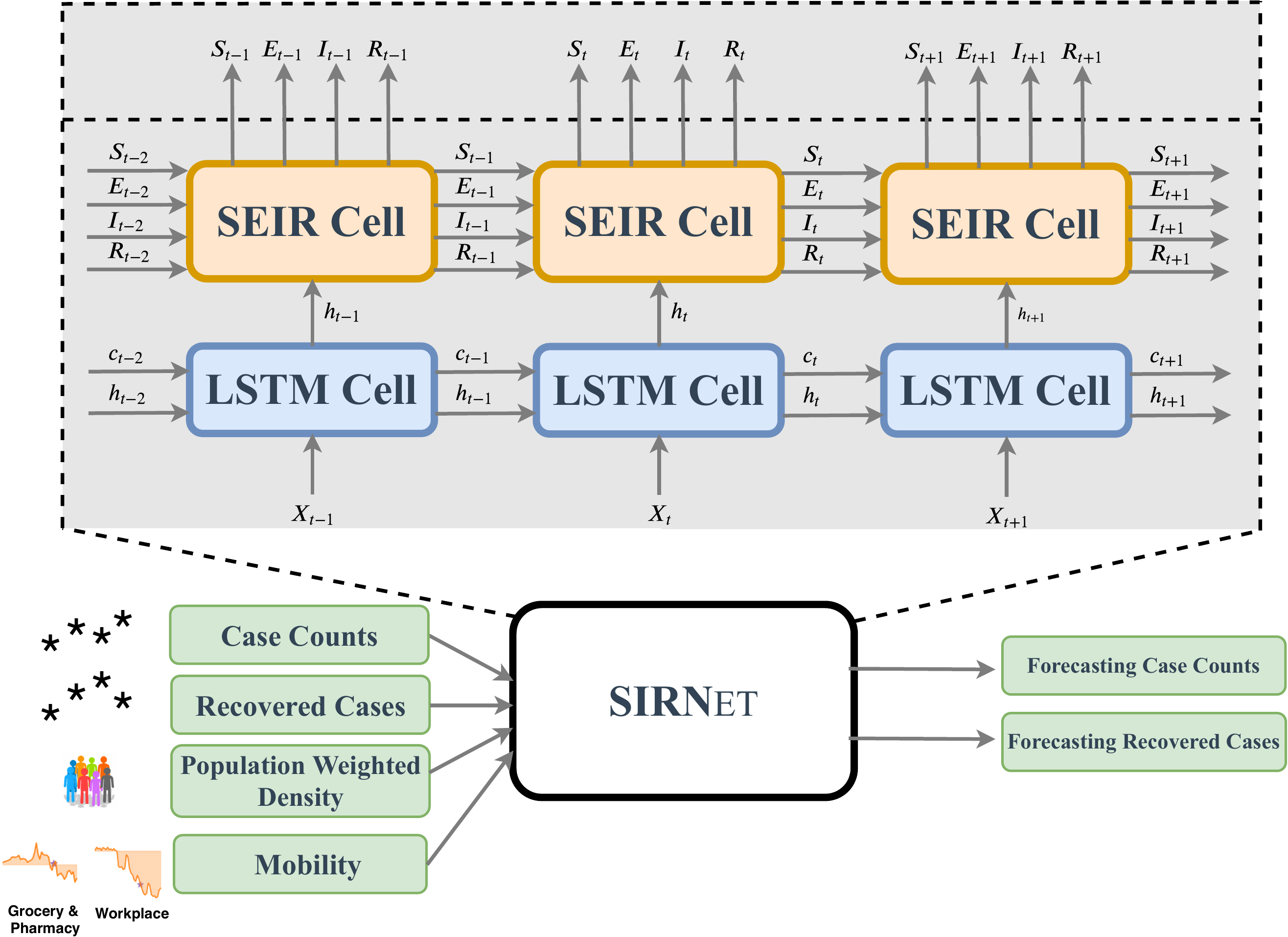}}
    \caption{High level visualization of the \EPINET{} architecture. (a) The RNN is a linear network with input layer $\in \mathbb{R} ^ 6$, hidden layer $\in \mathbb{R} ^ 4$, and output layer $\in \mathbb{R} ^ 1$, with $\mbox{ReLU}$ activation. $\mathbf{B_{i}}$ is the intractable contact rate. The RNN is a deep LSTM whose internal state is fed as input to the SEIR cell.}
    \label{fig:epinet}
\end{figure}

\subsection{Problem Formulation}
Here, we formalize the general form of this forecasting problem. We are given a set of time series that temporally enumerates active, recovered, and fatal cases of COVID-19. The data exhibits varying levels of geographical granularity, i.e., grouping by country, region, sub-region, etc., and irregular onset of the first case. Each sequence $\mathbf{Y} \in \mathbb{R} ^ {T \times 2}$ comprises $T$ timesteps, which varies between samples, each with the count of active cases and recovered cases. Provided this data, we desire to learn a model that is able to forecast future values of $\mathbf{Y}$. To aid in improving the fit, additional factors should be considered that may impact the infection, recovery, or mortality rates pertaining to the disease. We refer to these additional attributes as features which can be either scalar $\mathbf{x} \in \mathbb{R} ^ F$, spatial $\bm{\mathcal{X}} \in \mathbb{R} ^ {F_1 \times F_2 \times \dots \times F_n}$, and/or temporal $\mathbf{X} \in \mathbb{R} ^ {T \times F}$ ($\bm{\mathcal{X}} \in \mathbb{R} ^ {T \times F_1 \times \dots \times F_n}$ if spatiotemporal)%
\footnote{A tensor $\bm{\mathcal{T}}$ is a multilinear data structure (a higher-order matrix in layman's terms) and is denoted by bold uppercase calligraphic font.}.
With these definitions, the learning problem can be posed as follows (also see \eqref{equ: learning problem}). Given historical case data~$\mathbf{Y}$ and relevant attributes~$\bm{\mathcal{X}}$ for an area or multiple areas, can we model ($\mathcal{M}$) the latent trends of this data to forecast
future cases of COVID-19?
\begin{equation}
\begin{aligned}
& \underset{}{\text{minimize}}
& & \textup{cost}\left (Y_{t+k}, \mathcal{M}(\bm{\mathcal{X}}_t,Y_t; \theta_0, \theta_1, \dots)\right)%
\end{aligned}
\label{equ: learning problem}
\end{equation}

\subsection{Linear and LSTM Cells}
A high-level architecture visualization is shown in Figure~\ref{fig:epinet}. The historical time series and local raw data input can be fed to different types of Recurrent neural networks (RNNs) or a Linear cell. Linear cell can be a simple 3-layer network or a deep network with different activation functions. A rectified linear unit works best in most cases. RNNs have the ability to learn patterns from arbitrarily long spatiotemporal data, through cyclic connection of nodes in the network \cite{lipton2015critical}. An LSTM cell, a type of RNN, addresses some of the shortcomings in vanilla RNNs, such as vanishing or exploding parameters by adding memory \cite{schmidhuber1997long}. The LSTM is computed by \eqref{equ:LSTM-Topology} where $[W^{hI}_{ij}, W^I_{ij}, W^{hf}_{ij}, W^f_{ij}, W^{ho}_{ij}, W^o_{ij}, W^{hc}_{ij}, W^c_{ij}]$ are a set of weight tensors, $h^{t-1}_{j}$ vector represents the output of the hidden layer in the previous time step, $X_{i}$ is the input vector, $[B^{h}_j,B^{f}_j,B^{o}_j,B^{c}_j]$ represent bias vectors, and $\theta$ and $\Gamma$ are activation functions. 
\begin{equation}
\resizebox{0.65\textwidth}{!}{$
\begin{cases}
        I^{t}_j =\theta( B^{h}_j + \sum_{i=0}^{N_1} {h^{t-1}_{j} \times W^{hI}_{ij} + \sum_{i=0}^{N_2} X_{i} \times W^I_{ij}}),   \\
        f^{t}_j =\theta( B^{f}_j + \sum_{i=0}^{N_1} {h^{t-1}_{j} \times W^{hf}_{ij} + \sum_{i=0}^{N_2} X_{i} \times W^f_{ij}}), \\
        O^{t}_j =\theta( B^{o}_j + \sum_{i=0}^{N_1} {h^{t-1}_{j} \times W^{ho}_{ij} + \sum_{i=0}^{N_2} X_{i} \times W^o_{ij}}), \\
        \tilde{C}^{t}_j =\Gamma( B^{c}_j + \sum_{i=0}^{N_1} {h^{t-1}_{j} \times W^{hc}_{ij} + \sum_{i=0}^{N_2} X_{i} \times W^c_{ij}}), \\
        c^{t}_j = f^{t}_j \bigodot c^{t-1}_j + I^{t}_j \bigodot \tilde{C}^{t}_j, \\
        h^{t}_{j}= O^{t}_j  \bigodot \Gamma (c^{t}_j), \\
        y^{t}_j= \gamma( B^{y}_j + \sum_{i=0}^{N_3} {h^{t}_j \times W^y_{ij} } ) \\
\end{cases}$}
\label{equ:LSTM-Topology}
\end{equation}  

The \EPINET{} consists of a recurrent neural network to implement the temporal and population dynamics of an SEIR cell, and its framing as such allows us to introduce complex functions with learnable parameters, enabling mapping from salient input data (such as mobility) to the underlying properties of the epidemiological model.

\subsection{SEIR Cell}
One standard approach to epidemic modeling is compartmentalized models such as SEIR - with Susceptible $S$, Exposed $E$ (latent infected, but not yet infectious), Infected $I$, and Recovered $R$ (no longer infectious, also referred to as removed) states. The rate of change in these parameters is represented by the ordinary differential equations~\eqref{eq:seir_ds}-\eqref{eq:seir_dr} and parameterized by $\beta$ (effective contact rate/infectious rate learned from mobility data),
$\sigma$ (the incubation rate),
and $\gamma$ (recovery rate).
%
\begingroup
\allowdisplaybreaks
\begin{align}
    \label{eq:seir_ds}
    \frac{dS}{dt} &= -\beta S I
    \\[1ex]
    \label{eq:seir_de}
    \frac{dE}{dt} &= \beta S I - \sigma E
    \\[1ex]
    \label{eq:seir_di}
    \frac{dI}{dt} &= \sigma E - \gamma I
    \\[1ex]
    \label{eq:seir_dr}
    \frac{dR}{dt} &= \gamma I
\end{align}
\endgroup

The basic reproduction number representing the number of secondary infections from a primary individual in a completely susceptible population can be computed by,
\begin{equation}
    {R}_{0} = \frac{\beta}{\gamma}
\end{equation}

In the proposed \EPINET{} model, we use two different neural networks to learn the parameters $\beta$ and $\gamma$, based on population weighted density and cell-phone mobility data (latent information of the contact rate). The mobility data at country, state, and county level is used to predict the contact rate within those regions respectively. In the model, recovery rate can be treated as an individual trainable parameter, or it can be treated as a constant established by medical reporting.

In particular, the model attempts to learn $\beta(t)$ by mapping $\beta(x(t))$, where $x$ represents relevant temporal data (we consider only time steps of one day).  The SEIR cell's hidden states consist of the four compartmental groups normalized by population.

While our approach can be extended to many types of data, our work here is focused on one type in particular: mobility data.  Contact rate is a key parameter of the model and its modification through quarantine measures is an effective way to control the spread of the virus.  Contact rate is a function of population density as well as how people move and interact with each other. Traditional modeling can retrospectively estimate the change in contact rate brought about by policy changes (step-function changes), in our approach we build upon this technique to allow the integration of richer, daily information based on the actual activities of a population.  To this end, we begin with cell-phone based mobility information.

The mobility input vector, $x$, consists of mobility ratios (current mobility divided by nominal mobility) in 6 categories provided through \cite{COVID19C25:online}.  \EPINET's task is to use this feature vector to learn the resulting contact rate as a function of population mobility.  Through the use of the SEIR cell, we can map the output to case counts and learn the underlying mobility to contact rate function in an end-to-end fashion.

For the mobility model, the SEIR cell predicts contact rate according the following function:
\begin{equation}
    \beta(t) = {\textup{ReLU}(W \cdot x)}^p
\end{equation}

Our model is a linear combination of mobility contributions to an effective mobility, raised to the power $p$.  The parameterization of $p$ allows the model to learn the effective power of scaling mobility rather than simply assuming one, removing the need to justify a linear or quadratic relationship (our modeling exercises suggest the latter provides the best representation). The rectified linear unit function ensures that our model will only produce non-negative contact rates.


One of the primary challenges in modeling is that the underlying state is difficult to estimate.  The only data that is reliably available is the total case count, and the case count only represents a fraction of actual cases (with an estimated 50\%-80\% being underreported).  It also lags the true state of the system by several days.  Mobility data will drive exposure, exposure will drive the amount of the infectious, and the infections will in turn drive the number of cases.  To account for all of these factors, we use the 5-day incubation period \cite{TheIncub43:online,EarlyTra8:online,Clinical56:online} and add an additional 5 days to account for the delay between becoming infectious and receiving a positive test confirmation.  This delay in testing is not constant across time, nor is it consistent from location to location, but the measurable impacts of mobility on contact rate are most apparent when delay is taken into account.

We initialize the hidden state with the number of active cases at the onset of the epidemic, $I_0$.

\begin{figure}[H]
    \centering
    \includegraphics[width=.6\linewidth]{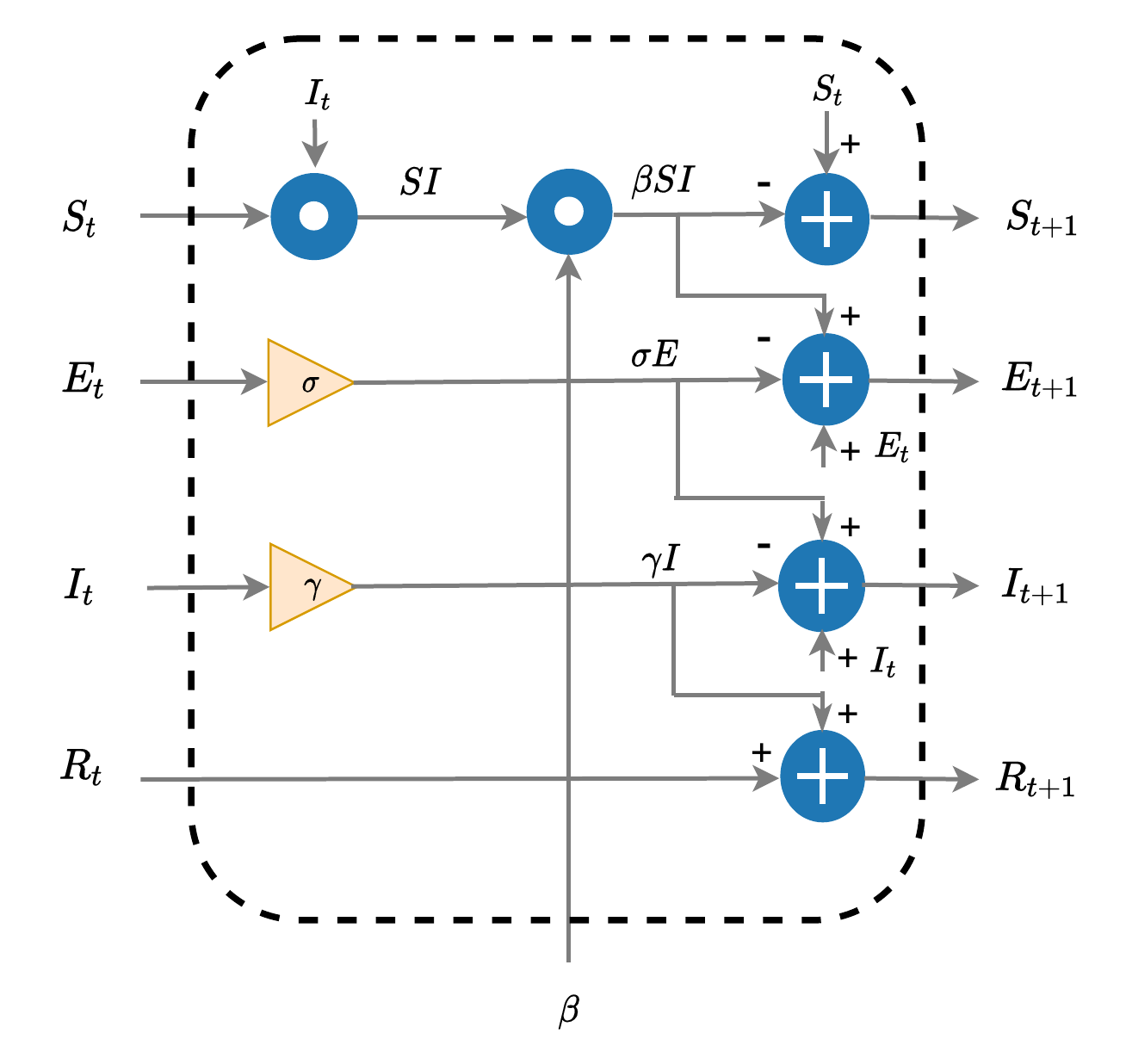}
    \caption{We frame the SEIR modeling as a recurrent neural network (RNN) architecture, introducing the SEIR-cell which encodes susceptible, exposed, infected, and recovered proportions in its hidden states. Open sourced on \textit{GitHub} -- download \textcolor{blue}{\href{https://github.com/Nu-AI/Livid-About-COVID}{here}}.}
    \label{fig:sirnet}

\end{figure}

\subsection{Assumptions}
The model uses case count resources from WHO~\cite{Situatio35:online}, CDC~\cite{Coronavi55:online}, European CDC~\cite{COVID1959:online}, NYTimes~\cite{nytimesc54:online}, and Texas DSHS~\cite{COVID19I60:online}. Specific data reported includes confirmed cases, deaths, and recovered cases. The population weighted density data and age-group data for US is obtained from the census website \cite{Populati65:online} and mobility data is captured from here \cite{COVID19C25:online} and here \cite{COVID-1954:online}. The assumptions made in the model are listed in Table \ref{table:Spec}. Additionally, the effective social distancing measures for each of the regions are populated from here \cite{COVID1912:online}. Testing and recovery data is collected from multiple sources for cross-validation, such as \cite{Coronavi87:online,CSSE-Cen85:online, COVID1912:online}.

\bgroup
\def\arraystretch{2}
\begin{table*}[ht!]
\caption{Parameters used in the models}
\label{table:Spec}
\centering
\begin{tabular}{@{}cccc@{}} 
    \toprule
    \hline
    \textbf{Parameter} 
    & \textbf{Real-world Value}  & \textbf{Experimental} &\textbf{References}  \\
  Incubation Period 
    & $\approx$ 5.2 days & -  & \cite{TheIncub43:online,EarlyTra8:online,Clinical56:online}   \\
    Viral Shedding &  $\approx$ 8-37  days  & & \cite{zhou2020clinical}\\
    Acute hospital beds & - & 40\% of total &\cite{CovidAct55:online}\\
    Testing Delays & $\approx$ 10 days & & \cite{Whysomec64:online}\\
    Initial infected & First reported case & -&\\
     \hline
    \bottomrule
\end{tabular}
\vspace{-0.5mm}
\end{table*}
\egroup

\section{Results and Analysis}
\subsection*{Model analysis for different geographical regions:} 
\begin{figure}[H]
    \centering
    \includegraphics[width=\linewidth]{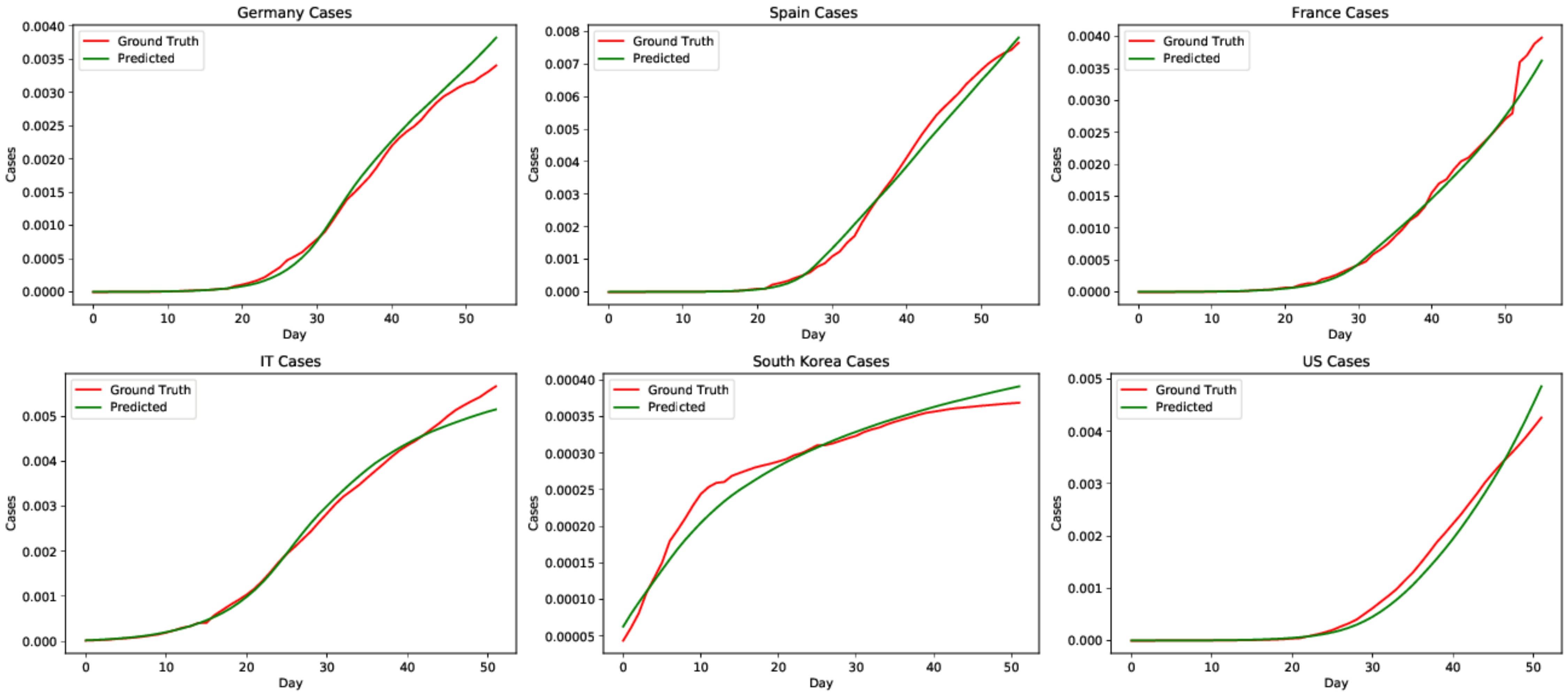}
    \caption{\EPINET{} predictions on previous and current CDC reported cases data for the United States, Italy, Spain, Germany, France, and South Korea.}
    \label{fig:comp_final_countries}
\end{figure}

\begin{figure}[H]
    \centering
    \includegraphics[width=\linewidth]{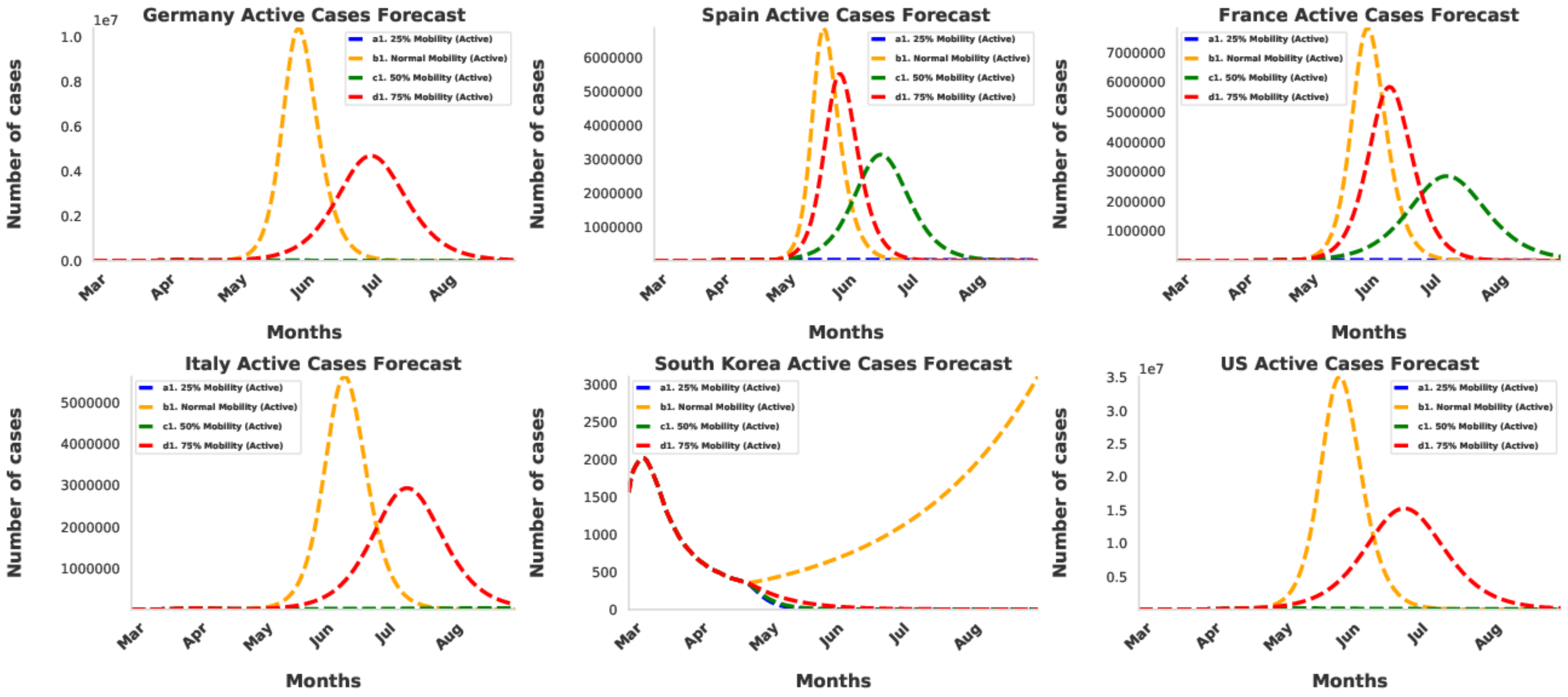}
    \caption{\EPINET{} forecast of active cases for the United States, Italy, Spain, Germany, France, and South Korea.}
    \label{fig:active_final_countries}
\end{figure}

\begin{figure}[H]
    \centering
    \includegraphics[width=\linewidth]{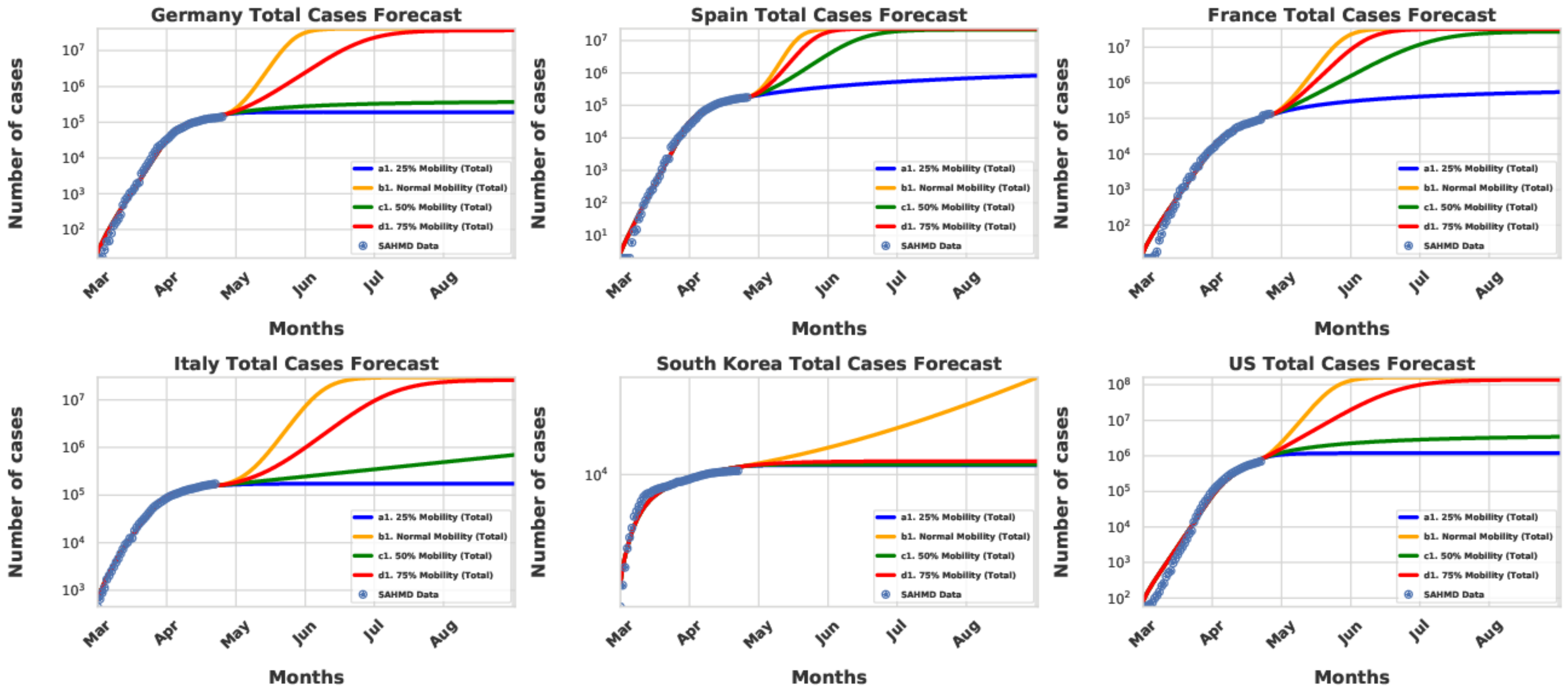}
    \caption{\EPINET{} forecast of total cases for the United States, Italy, Spain, Germany, France, and South Korea.}
    \label{fig:total_final_countries}
\end{figure}

\begin{figure}[H]

    \centering
    \includegraphics[width=1.0\textwidth,height=0.7\textheight,keepaspectratio]{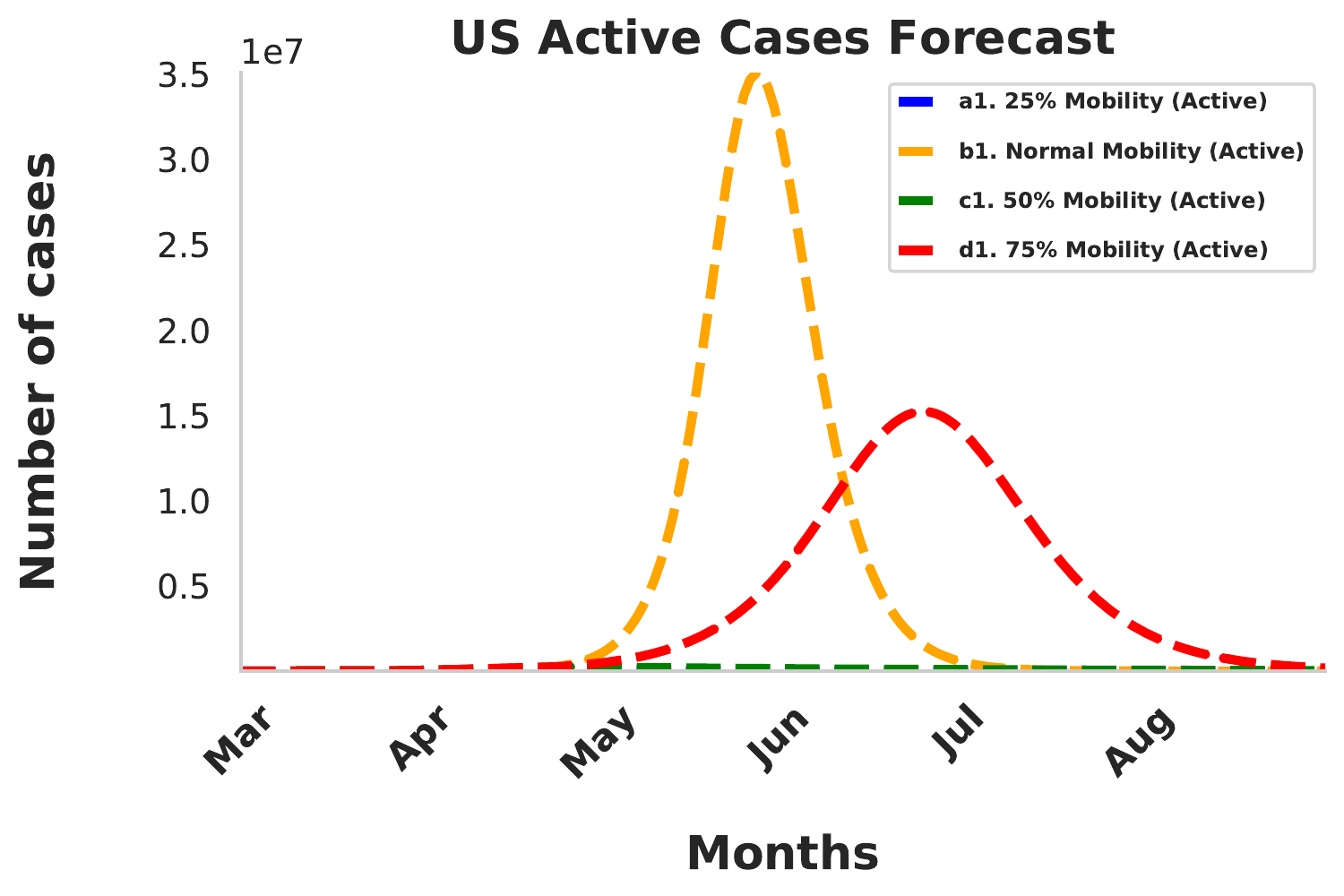}
   \begin{picture}(0,0)
\put(-140,190){\adjincludegraphics[height=3.1cm, trim={2cm 0 0 0},clip]{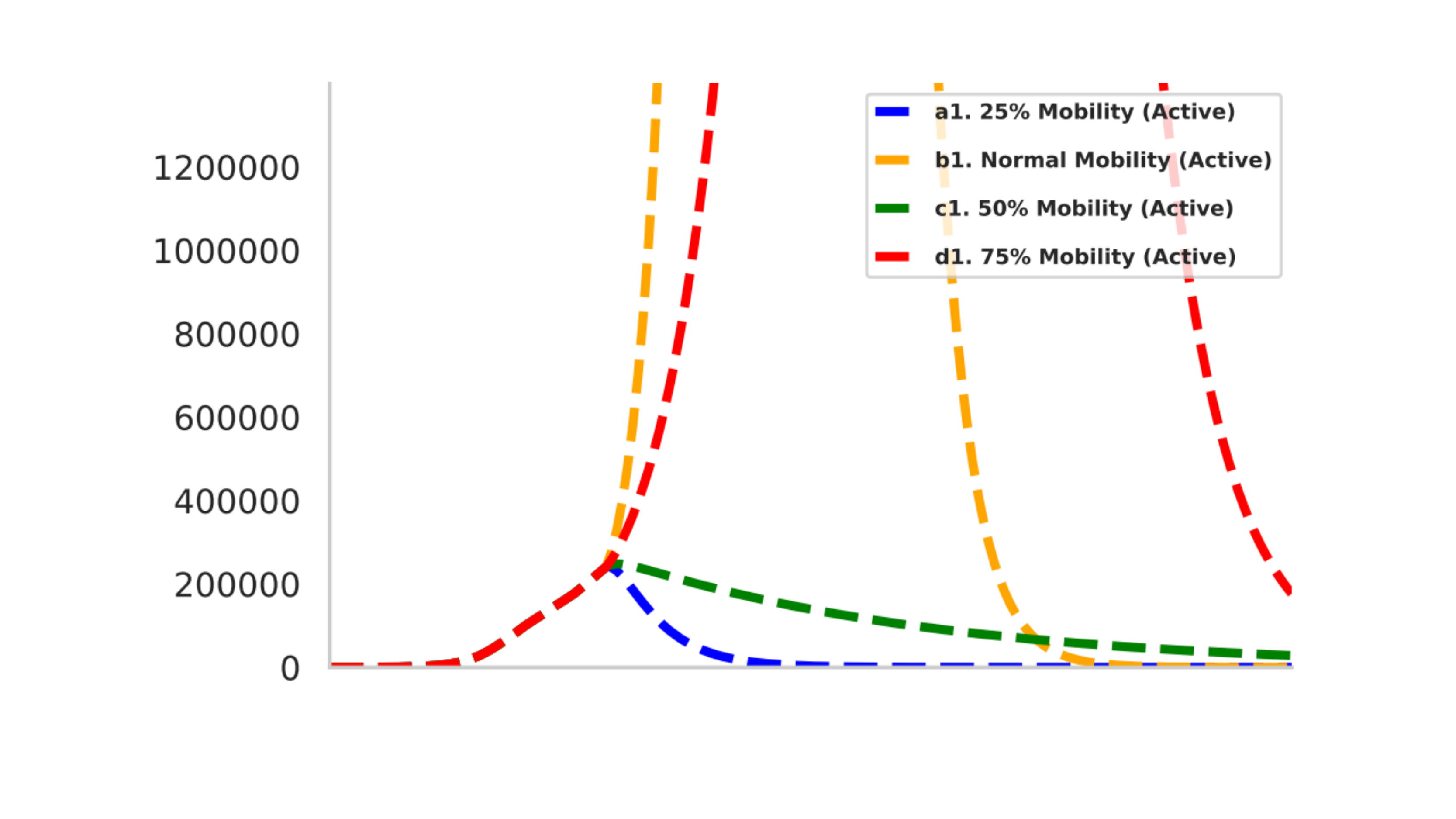}}
\end{picture}
\caption{Forecasting of active cases for USA under four different scenarios of physical distancing. Mobility data is real-time cell phone/mobile device location for USA collected from here \cite{COVID19C25:online}. Inset on the top left is further magnified onto the active cases with mobility  of 25\% and 50\%. The scale is reflective of this change.}
    \label{fig:active_zoom_us}
\end{figure}

\begin{figure}[H]
    \centering
    \includegraphics[height=.85\textheight]{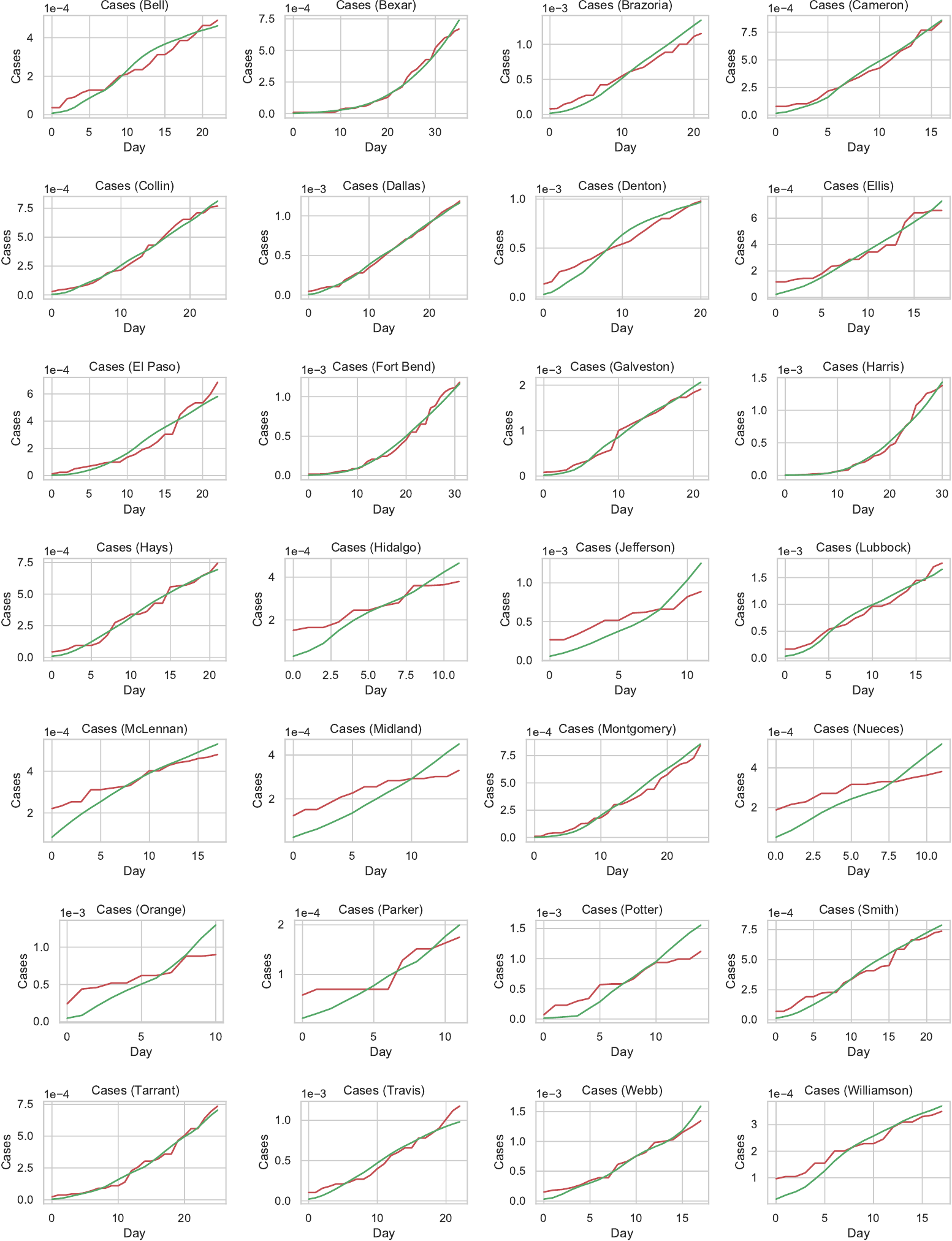}
    \caption{Model fit across the top 28 counties in TX, with all six mobility data points available. The physical distancing scenario is trained on 51 regions around the world encapsulating different levels of intervention. The red line is the ground truth total case data and the green line is the predicted total case data.}
    \label{fig:TX_county}
\end{figure}

\begin{figure}[H]
    \centering
    \includegraphics[width = 5in]{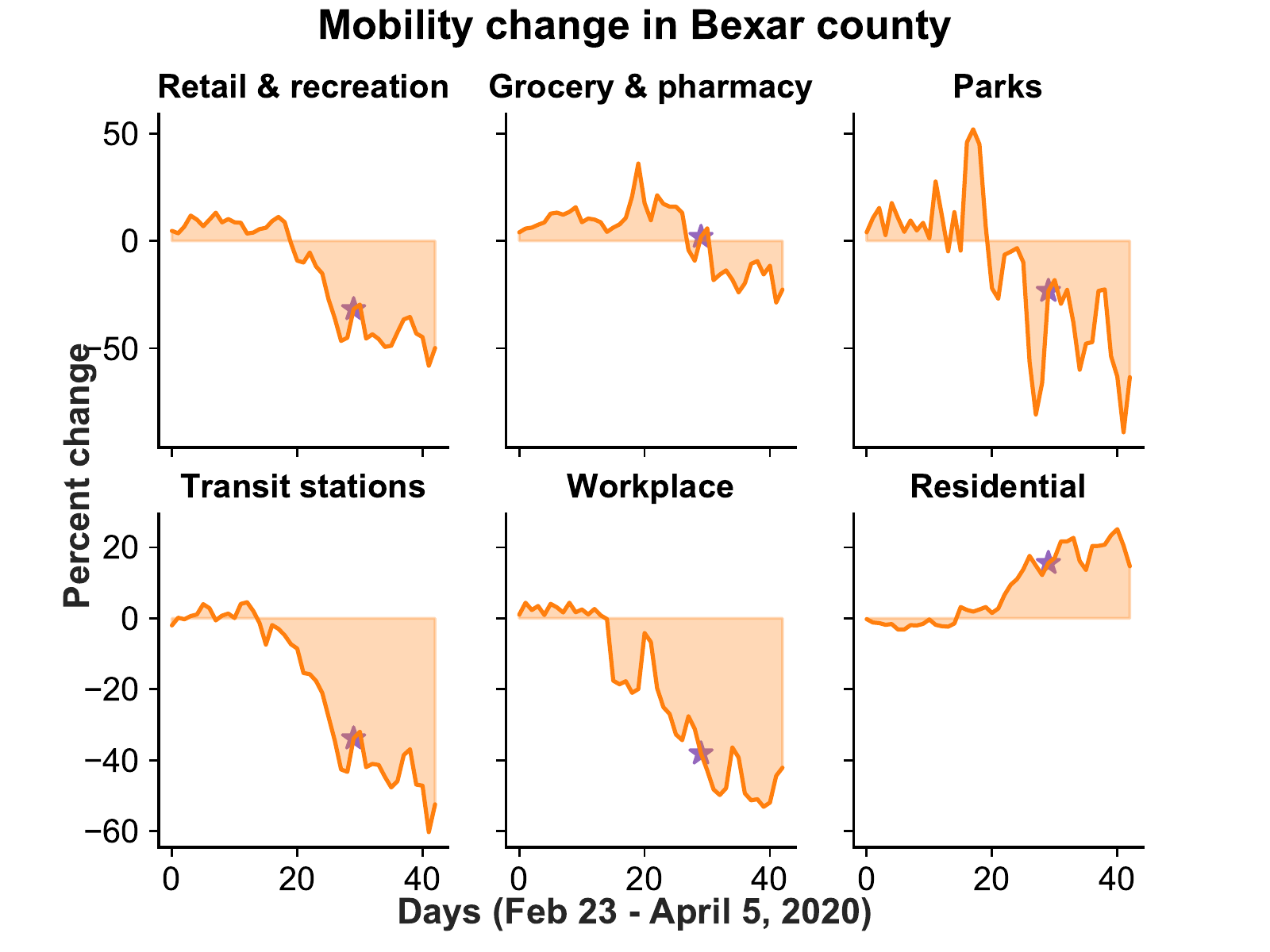}
    \vspace{-2mm}
    \caption{Sample mobility data changes in Bexar County, Texas from Feb 23\textsuperscript{rd} to April 11\textsuperscript{th}. The star indicates when social distancing measures were implemented in the county (March 23\textsuperscript{rd}). Data is collected using webscraping tools developed by the team. By comparison, mobility in Italy has decreased by $\approx$ 70\% to non-essential locations, with the lockdown procedures. Data Source: Google LLC \cite{COVID19I60:online}} 
    \label{fig:Bexar_Mobility}
\end{figure}

The \EPINET{} was evaluated on different geographical regions. Figure~\ref{fig:comp_final_countries} shows the fit for predicting total cases in a region based on mobility data compared to the ground truth in several countries. Figures~\ref{fig:active_final_countries} and~\ref{fig:total_final_countries} show the models forecast for different mobility levels in each country for active cases and total cases respectively. These figures demonstrate that the proposed \EPINET{} is able to fit the case count by region well using mobility information to determine the contact rate of an SEIR cell. Based on the projected forecasts, we observe that a continuation of quarantine level mobility will result low case counts. If the mobility restrictions are reduced to 50\% nominal mobility, the model shows that this is near the edge of stable peak cases where in some scenarios the curve stays at a low peak, while for others the peak increases drastically compared to the quarantined mobility and occurs much later. The third scenario is 75\% nominal mobility which based on the model is expected to result in a slightly delayed peak approximately 2/3 the maximum peak during normal mobility. The exception is South Korea, where at this stage even a return to 75\% mobility is not expected to result in a second wave. A zoomed in figure of the forecast of active cases in the US is shown in Figure~\ref{fig:active_zoom_us} to better illustrate the difference between mobility levels. In general, these results suggest a continuation of quarantine level mobility or at least below 50\% nominal mobility for the immediate future. Figure~\ref{fig:TX_county} shows the model fit with ground truth for the top 28 counties in Texas. In general the model fits well where the case counts are higher than 50 and as data becomes richer, the fit improves significantly. Figure~\ref{fig:Bexar_Mobility} highlights a sample of the mobility trends used in all our simulations. \textbf{It is important to note that this data reflects a sample space of mobility for the region and might be missing information on key populations that do not use specific types of devices. Adding finer granular information and data from multiple data providers can alleviate this concern, as it becomes open-sourced.}

\clearpage  



\begin{figure}[H]
    \centering
    \includegraphics[width = 6in]{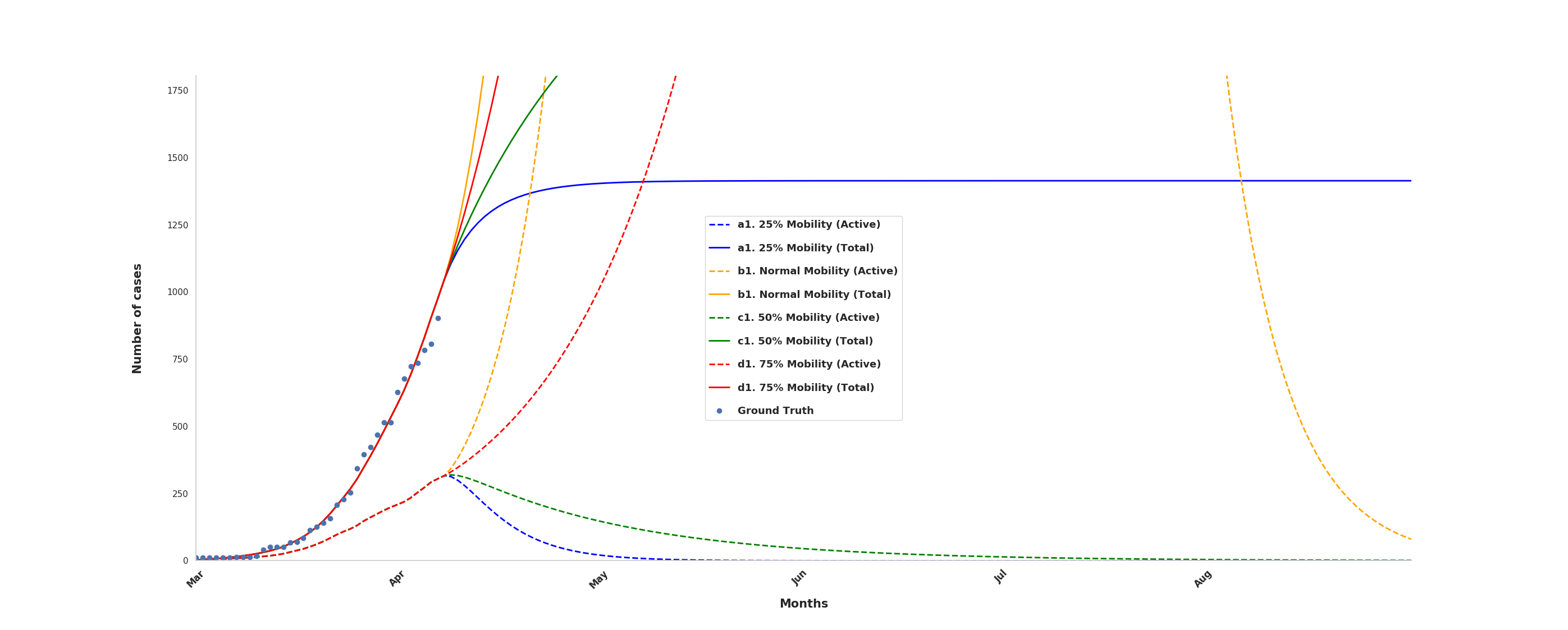}
    \vspace{-2mm}
    \caption{Forecast for Bexar County, TX.  All models are driven initially by recorded mobility and then split to forecast different scenarios.  While current mobility (as of 11 April 2020) is approximately 50\% of nominal, relaxing measures at this time could result in runaway growth. } 
    \label{fig:Bexar_Forecast}
\end{figure}

\begin{figure}[H]
    \centering
    \includegraphics[width = 6in]{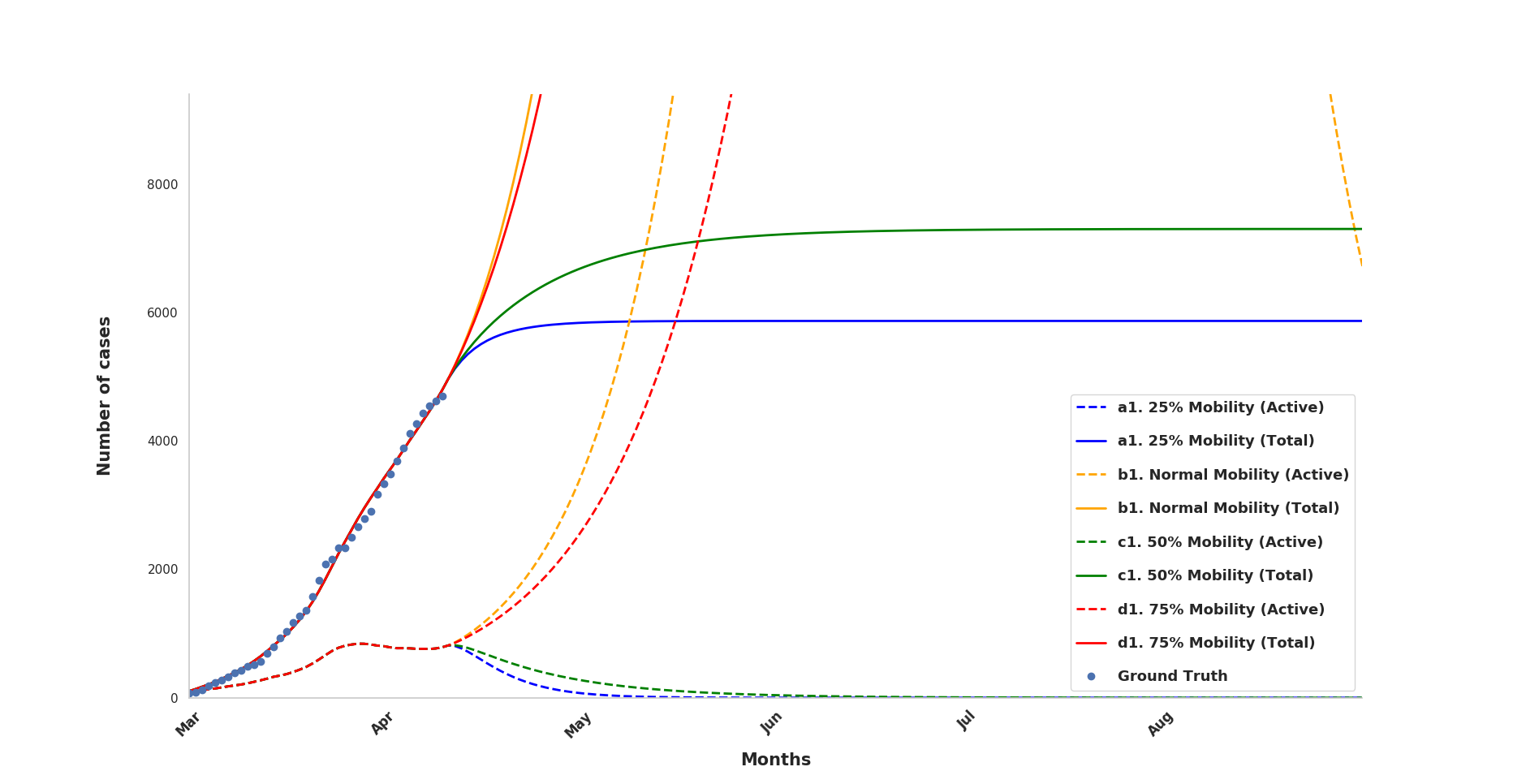}
    \vspace{-2mm}
    \caption{Forecast for King County, WA, where a longer period of infection enables better model fitting.  King County's heavy mobility restrictions are clearly dropping the $R_{0}$ below 1.}
    \label{fig:King_Forecast}
\end{figure}

\subsection*{Why Does Reduced Mobility Help?}
Our model shows that the epidemic is highly sensitive to changes in mobility rates. Figure \ref{fig:Bexar_Forecast} and Figure \ref{fig:King_Forecast} shows examples of two counties at different peak points of spread and how changing mobility rate is impacting these counties (Bexar county in Texas and King County in WA). 
SEIR cell dynamics show the critical points in reproduction number, in which we project mobility levels showing three possible outcomes.  First, heavy mobility restrictions can drive the contact rate and reproduction number down so that the virus is eliminated from the population.  The number of active cases and the number of total cases are kept to minimal levels, though the population remains susceptible to second waves.  Second, at higher levels of mobility (those approaching nominal), the virus reproduces prolifically in the population, with exponential growth ending as herd immunity is reached.  This results in both overwhelmed medical resources due to active cases and a large number of deaths due to the total case count.  A third possibility is a reproduction number of approximately 1.  In this scenario, active case count remains low; however, the virus would continue to work its way through the population until herd immunity is reached, resulting in a large number of total cases and deaths as a slow tail gradually tapers off. \textbf{In the mobility scenarios tested across countries or counties, mobility $>$ .7 leads to an uncontained outbreak, mobility $<$ .5 results in a local elimination of the virus, and those in between having slower peaks.}

Remobilizing a population will require careful monitoring to ensure that the critical reproduction number is not approached.  It should also be considered that while our model introduces a new feature for relating human activity to mobility, it is certainly not the only factor to consider.  Changes in behavior, including social distancing and hygiene habits, are certainly also contributing to the reduced reproduction rate in a way that cannot be considered independently (their effects are reflected in mobility data, but are difficult to predict in future mobilization).  It is imperative that policymakers consider the dynamic nature of infection. \textbf{The current projected peak is the worst time to begin relaxing quarantine measures. Also, there is only one peak for current mobility because of the drastic measures that have been taken thus far}.  Continuous monitoring and modeling will be indispensable tools for containing the outbreak, and models will continue to improve as more data becomes available.

\begin{figure}[H]
\centering
\includegraphics[width=0.5 \linewidth]{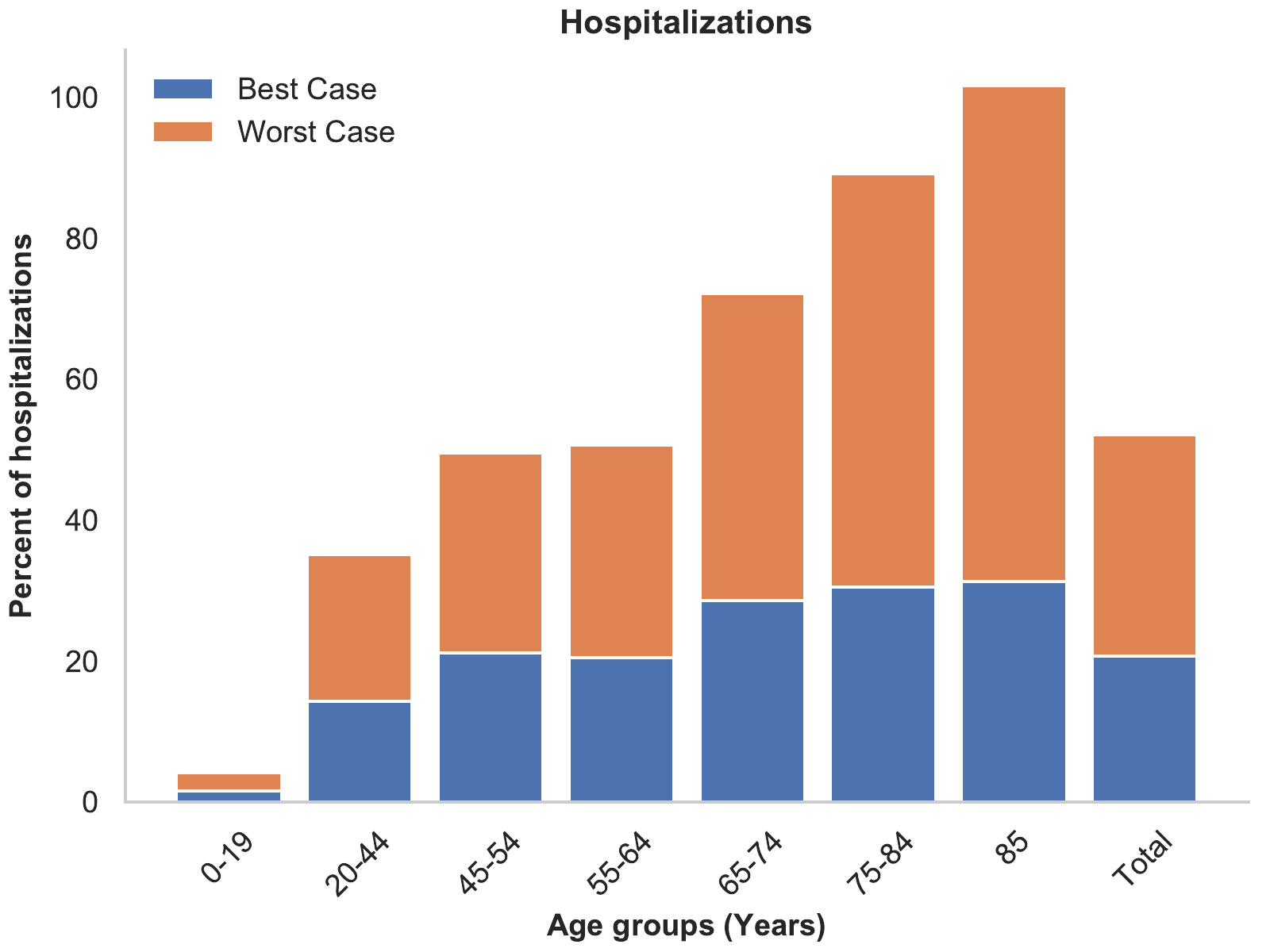}
\vspace{-2mm}
    \caption{Hospitalization rate for SARS-CoV-2 in a sample county (Bexar, TX) based on age group. The population by age group data for 2018 is obtained from here \cite{Populati65:online}.} 
    \label{fig:AI_Bexar_hosp}
\end{figure}

Based on the projections for a region, the \EPINET{} model can calculate the potential hospitalization rates by age group, for best and worst case scenarios. This is computed based on local age demographics, forecasted active cases, and estimated hospitalization rates by age group from experimental data. An example scenario for a county is shown in Figure~\ref{fig:AI_Bexar_hosp}.

\subsection*{Error Tolerance} When using any ML or statistical models to forecast trends, it is important to consider the confidence interval or margin of error for the predictions. \EPINET{} is currently trained on a specific region it is forecasting, with region-specific assumptions about under-reporting, delay in reporting, the recovery rate, and the transition rate from exposed to infected. In future work it will be necessary to account for the error range for each of these variables based on global reported data, and use this to predict the potential fluctuation in forecasted scenarios. Another important extension to \EPINET{} is to extend learning to multiple regions, providing a more generalized forecast that can capture distinctions between different regions.

\section*{Conclusions and Discussion}

Our work takes a multidisciplinary approach to address modeling the spread of COVID-19. \EPINET{} is a hybrid between epidemic modeling, physical science, and machine learning. The benefit of epidemic modeling, is constraining our network to produce meaningful variables from a physical standpoint which adds an intuitive understanding of how the model is forecasting and provides an approach for overcoming limited or missing real-world data samples. On the other hand, machine learning provides a tool for translating variables, such as mobility, non-pharmaceutical intervention, and population demographics, into variables that impact an epidemic model. It also allows us to discover relationships between real-world trends and the impact on the spread of COVID-19, as well as model scenarios such as relaxing social-distancing policies. We believe both components are necessary to develop an insightful model to aid in understanding the impact of non-pharmaceutical interventions on COVID-19.

Similar to other approaches, we base our study on several biologically observed data and real-world datasets. We demonstrate how new tools can be created to better exploit available quantitative measures in the fight against COVID-19.  By integrating reliable metrics and well-studied infection dynamics, we create an approach that is deeply data-driven and science-based.  Our studies confirm the effectiveness of reduced mobility for limiting the reach of the pandemic, and our models provide a means of forecasting the effects of different mobility scenarios. 

\EPINET{} is in an early iteration and requires extensive sensitivity analysis to understand the range of impact of different parameters. Additionally, exhaustive mobility data combined with non-pharmacological intervention datasets can improve the network predictions. Since several datasets are proprietary and limited by data user agreements, it will be important to establish good data collection and standardization practices to address catastrophic events.

Given the substantial risk of reintroduction of the SARS-CoV-2, it is critical to reinforce balanced social distancing measures in the coming months to reduce the impact on the healthcare system, general public, and economic prosperity. Resource limitations in a rapidly growing pandemic demand compelling resource utilization choices. Of importance is to note that the data-driven AI models provide a window into understanding the potential impact and should be treated as a qualitative guidance due to the rapid changes associated with the data collection, testing strategies, reporting, and the virus transmission.

\section*{Acknowledgements}
\noindent Authors would like to thank all the open-source data providers, which was critical in timely analysis of the spread. We are grateful for researchers S. Hamed Fatemi Langroudi, Pankil Shah (UTHSCSA), and Tej Pandit who have helped in the simulations and tool flows. The authors would like to acknowledge the research organizations and their leads at UTSA, UTHSCSA, and SwRI for their thought leadership and  open support for multidisciplinary collaboration. Research team also recognizes several members of the community in the city of San Antonio and UT System who have aided in data access and critical discussions in understanding the virus. We appreciate members of Project Alpha, Amina Qutub and Hongjie Xie, who continue to share insights on COVID-19 recovery and tracking tools.

\paragraph{Conflicts of Interest:} The authors declare that they have no conflicts of interest to report
regarding the present study.
\bibliographystyle{style/Science}
\bibliography{ref}

\section*{Supplementary materials}

\begin{figure}[H]
    \centering
    \includegraphics[width = 4in]{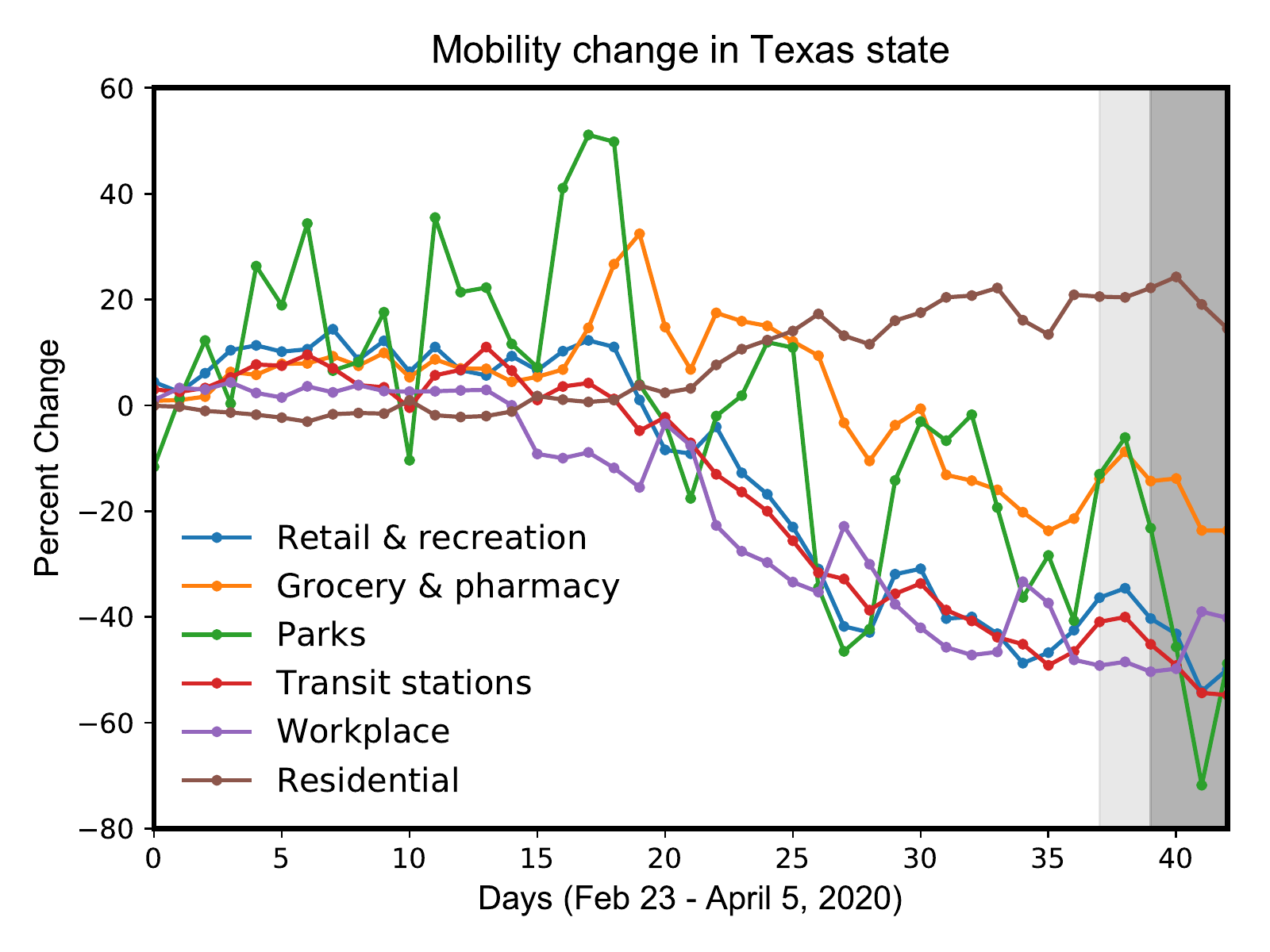}
    \vspace{-2mm}
    \caption{Mobility changes in the state of Texas over the past 6 weeks. The most recent datapoint is April 5\textsuperscript{th}. Data is collected using webscraping tools developed by the team. Mobility to workplace and residential areas has decreased $\sim$\,42\%, over the past 40 days. The shaded regions represent the days when stay at home order was announced (March 31\textsuperscript{st}) and was effective (April 2\textsuperscript{nd}) in Texas \cite{stay_at_home:online}. Data Source: Google LLC\cite{COVID19C25:online}} 
    \label{fig:TX_Mobility}
\end{figure}

\end{document}